# Laboratory analogues simulating Titan's atmospheric aerosols: compared chemical compositions of grains and thin films

Nathalie Carrasco[1,2], François Jomard[3], Jackie Vigneron[4], Arnaud Etcheberry[4], Guy Cernogora[1]

[1] LATMOS, Université Versailles St Quentin, UPMC Univ. Paris 06, CNRS, 11 blvd d'Alembert, 78280 Guyancourt, France
[2] Institut Universitaire de France, 103 bd Saint-Michel, 75005 Paris, France
[3] GEMAC, Université Versailles St Quentin, CNRS, 45 Avenue des Etats Unis, FR78035 Versailles, France
[4] ILV, Université Versailles St Quentin, CNRS, 45 Avenue des Etats Unis, FR78035 Versailles, France

Corresponding author: E-mail: nathalie.carrasco@latmos.ipsl.fr

## Abstract

Two sorts of solid organic samples can be produced in laboratory experiments simulating Titan's atmospheric reactivity: grains in the volume and thin films on the reactor walls. We expect that grains are more representative of Titan's atmospheric aerosols, but films are used to provide optical indices for radiative models of Titan's atmosphere.

The aim of the present study is to address if these two sorts of analogues are chemically equivalent or not, when produced in the same $N_2$-$CH_4$ plasma discharge. The chemical compositions of both these materials are measured by using elemental analysis, XPS analysis and Secondary Ion Mass Spectrometry. The main parameter probed is the $CH_4/N_2$ ratio to explore various possible chemical regimes. We find that films are homogeneous but significantly less rich in nitrogen and hydrogen than grains produced in the same experimental conditions. This surprising difference in their chemical compositions could be explained by the efficient etching occurring on the films, which stay in the discharge during the whole plasma duration, whereas the grains are ejected after a few minutes. The higher nitrogen content in the grains possibly involves a higher optical absorption than the one measured on the films, with a possible impact on Titan's radiative models.

# Introduction

Titan, the largest satellite of Saturn, is surrounded by a brownish opaque photochemical smog. This smog is made of solid organic aggregates of a few hundreds of nanometers, rich in nitrogen, and with a global $C_xH_yN_z$ composition [1-3]. The chemical complexity of this smog is far from being understood despite the numerous *in situ* data by the ongoing Cassini-Huygens mission.[1, 4-6] Several experiments have been developed to infer their chemical composition by studying laboratory analogues.[7-9] According to the experimental setups, the analogues (also named tholins in the following) are produced on the wall of the reactor, or in the volume, producing respectively organic thin films or spherical shaped individual grains of a few thousands nanometers in diameter.[10, 11]

Little attention has been drawn to the possible effect of such different growth pathways on the chemical composition of the analogues. Individual grains and thin films are usually considered as equivalent materials in terms of their chemical signatures, and used alternatively depending on the involved analysis technique. Ellipsometry for example requires exclusively thin films to determine the optical indices of the material [12, 13], whereas high-resolution mass spectrometry analysis has been performed on grains only.[14, 15] However an infrared spectroscopic comparative study showed some differences between grains and thin films produced in the same experimental conditions. [16]

The aim of this work is to address this common paradigm by comparing the chemical composition of both solid materials produced simultaneously in a plasma experiment simulating the formation of solid organic aerosols in the atmosphere of Titan.[17, 18] The chemical composition of these tholins grains has been extensively studied in the past by infrared absorption spectroscopy, mass spectrometry and elemental analysis.[16, 19, 20] The chemical composition of the tholins thin films has to date only been globally characterized by infrared absorption spectroscopy[21]. The present work provides the first chemical analysis of tholins thin films along their depth profile, using XPS (X-ray Photoemission Spectroscopy) and SIMS (Secondary Ion Mass Spectrometry) analysis. Those two techniques are commonly used for organic films analysis in general,[22-25] and are well adapted for nitrogen containing films in particular[26-28]. Moreover two kinds of substrates are used, made of $CaF_2$ and $SiO_2$, to analyze the sensitivity of the solid deposition to the nature of the substrate.

# Experimental Section

## Sample production

Experiments are carried out with the PAMPRE set-up, based on a Radio-Frequency Capacitively Coupled Plasma (RF CCP) at 13.56 MHz.[18] RF plasmas are also used in nitrogen methane mixtures for the production of CNx hard films [29] or hydrogenated amorphous carbon films (a-C:H) [30]. Their efficiency to produce solid material explains their interest to simulate the formation of Titan's aerosols.

The plasma discharge is produced between a polarized cathode and a grounded metallic grid (anode) surrounding and confining the plasma. The same power discharge of 30 W is injected in the whole electric circuit for all the experiments. Half of the injected power is effectively consumed to produce the discharge itself.[31] The gas mixture is continuously injected in the reactor at 55 sccm and pumped through a rotary valve vacuum pump. The gas flow composition is adjusted in order to introduce from 0% to 10% of $CH_4$ in $N_2$. source for the plasma chemistry. Two high purity gas bottles are used, one of pure $N_2$ and one containing a $N_2$–$CH_4$ mixture with 10% $CH_4$ [20]. Four mixtures conditions are chosen here to scan chemical conditions in agreement with the methane concentration profile in the atmosphere of Titan [32]: 1, 2, 5 and 10 % of $CH_4$ diluted in $N_2$. Experiments are performed at room temperature and a 0.9mbar pressure. The gas in the reactor is slightly warmed by the plasma discharge reaching a stable temperature of 315 and 340 K for methane amounts of 2 and 10% respectively [17]. The reactor walls are slowly warmed accordingly but no change is observed in the 2-hrs experiments performed here.

RF CCP discharges are well-known for producing simultaneously thin films on substrates and solid particles (grains) in the volume of reactive gas mixtures.[33] In our experimental set up, the thin organic films are produced on the grounded anode, whereas levitating grains are produced simultaneously in the gas volume without interaction with the walls.[34] To collect the films on the grounded electrode, two optical disks with a 1 cm diameter and a 1 mm thickness made of respectively $CaF_2$ and $SiO_2$ are symmetrically placed in the reactor to provide comparable substrates for thin film deposition. These two substrate materials are chosen to be chemically different but with similar geometry and similar electric properties in the plasma discharge. A 2 h experiment duration is chosen, to limit the thickness of the films below 1 μm.[12]

The film synthesis is a reproducible process. In previous studies we chose the same experimental conditions: silicon substrates, 5% of injected $CH_4$, 55sccm gas flow, 30W of injected power [12, 21]. Gautier et al (2012) synthesized three films in a 3h-duration experiment, and Mahjoub et al (2012) synthesized a film in a 2h-duration experiment. The obtained thicknesses are reported in *Table 1*. The growth rates are in agreement for the four samples showing that the films grow linearly, at least in the few-hours duration range of the experiments, and are well reproducible.

| Synthesis duration | Thickness | Thickness/h | Reference |
| --- | --- | --- | --- |
| 3h | 1250 ± 40 | 417 ± 13 | [21] |
| 3h | 1340 ± 40 | 447 ± 13 | [21] |
| 3h | 1300 ± 40 | 433 ± 13 | [21] |
| 2h | 850 ± 30 | 425 ± 15 | [12] |

*Table 1: Films thicknesses obtained in various synthesis, using the same experimental conditions: Silicon substrates, 5% of injected $CH_4$, 55 sccm gas flow, 30 W of injected power. Three substrates were placed in the plasma box for the same experiment in [21]. The uncertainty of the film thickness is given by the non-uniformity calculated by Complete-EASE™ software [12].*

The films samples are carefully collected after air exposure of the reactor and stored in individual boxes. The syntheses are performed in a one single experimental campaign of a few weeks; just before the analysis of the films in order to limit as much as possible the time interval between the synthesis and the analysis. Films are analyzed here by X-ray Photoelectron Spectroscopy (XPS) and Secondary Ion Mass Spectrometry (SIMS). And solid grains have been previously analyzed by elemental analysis in [20].

**X-ray Photoelectron Spectroscopy (XPS)**

The first step of the analysis is a chemical probe by XPS along the depth of the films. This is achieved using a Thermo Scientific K-Alpha instrument, with a monochromatic Al-K α X-Ray source (1486.6 eV) located in the ILV laboratory. Charge compensation is used to overcome charging effects on the organic films. The X-ray spot is set at a 400 × 400 μm size on the substrate. The pass energy of the analyzer is used in Constant Analyzer Energy mode, with a 50.0 eV pass. The energy step size is set at 0.1 eV. Sputter depth profiling is performed using an Argon ion gun (2 keV energy, 10 μA current), with an etch time of 10 s. An etch time of 10 s is equivalent to an etching thickness of 15 nm, calibrated using $Ta_2O_5$. After each etching step, XPS measurements are performed. The data treatment is performed using a Shirley baseline and a 30% Lorentz-Gaussian distribution for the peak fitting. XPS measurement tracks every component of the organic film but hydrogen.

The elemental composition is therefore given as normalized results, taking into account the expected elements composing the films and possible contributions of the substrate (Ca and F for films deposited on $CaF_2$ substrates; and Si and O for $SiO_2$ substrates), but no hydrogen. Therefore XPS analysis directly provides the C/N ratio of the films, but the hydrogen content, important for characterizing an organic material can unfortunately not be given by this only technique.

**Secondary Ion Mass Spectrometry (SIMS)**

After XPS analysis, the organic solid film is covered with a thin gold film for SIMS analyses. Those are performed using a Cameca IMS4F ion microscope located in the GEMAC laboratory. The organic film is bombarded by a 14.5 keV $Cs^+$ ion beam, of 10 μm diameter. The etch time can reach 3000 s.

The sputtered ions are collected in negative mode, preventing any nitrogen detection. The ions monitored are therefore $H^-$, $^{12}C^-$, $^{16}O^-$ as film constituents, but also $^{30}Si^-$ and $^{19}F^-$ as substrate constituents. In the case of $SiO_2$ substrates, the oxygen detected is the cumulated signature of the film and the substrate. This collection mode is stable and useful for qualitative purpose, but it does not allow absolute quantitative analysis. In the following it will be used to quantify the evolution of the C, N and H concentrations in the films in arbitrary units.

To provide a quantitative information, an additional analysis is performed by SIMS on the 10% thin film sample, monitoring in positive mode the molecular ions M-Cs$^+$ (M is a neutral atom extracted from the film). The energy of the bombardment is settled at 5.5 keV in this case. Molecular ions are known to be more representative of the elemental composition of the sample than the analysis of the direct sputtering ions.[35, 36] This quantification is validated here on the case of the SiO$_2$ substrate *Figure 1*: when the SIMS Cesium ion beam reaches the substrate, a ratio of 1.9 is found between the molecular signature of O-Cs$^+$ and Si-Cs$^+$ instead of the 2 value expected.

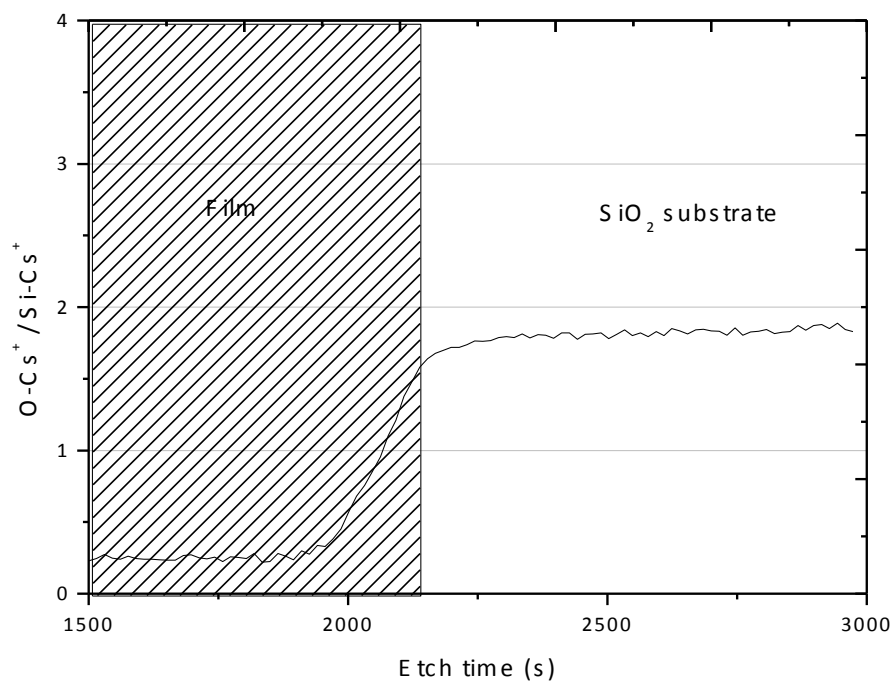

*Figure 1: Intensity ratio of the molecular ions O-Cs$^+$/ Si-Cs$^+$ detected by SIMS on the film sample with a 10% methane gas mixture deposited on a SiO$_2$ substrate. The shadowed area on the left corresponds to the organic film layer, and the clear area on the right to the SiO$_2$ substrate.*

# Results: XPS and SIMS analysis of the thin films

## XPS measurements: homogeneity of the films along thickness-profiles and existence of an oxidized surface-layer

The homogeneity of the films is probed by XPS profiling after argon ion sputtering on the first hundreds nanometers. Profiles obtained by SIMS and XPS on a $SiO_2$ substrate, with a 10% $CH_4$ condition are given on **Figure 2**. The very low amount of Si (< 0.1 %) measured by XPS confirms the absence of substrate contamination in the film.

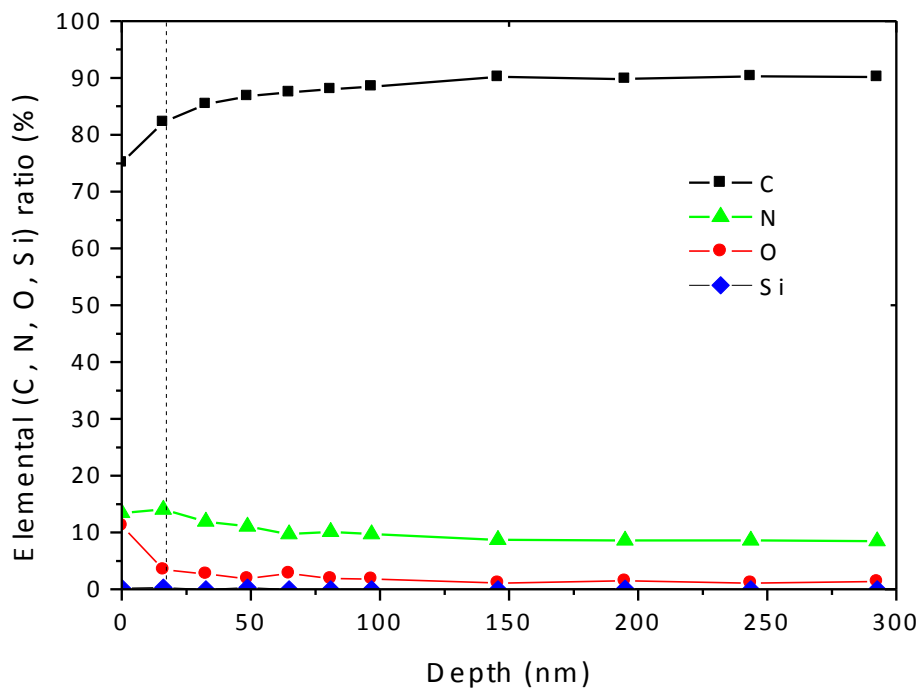

*Figure 2: XPS profile of a thin film with a 10% methane gas mixture on $SiO_2$ substrate. The dashed vertical line presents the limit of the oxidized superficial layer.*

The oxygen contribution is quantitatively given by the XPS data: oxygen counts for 1% inside the films. As the real elemental composition is even lower because of the hydrogen contribution unseen by XPS, tholin films can be considered as almost oxygen-free. This confirms the absence of water adsorbed or air traces in the plasma reactor during the synthesis, and that no diffusing oxidation occurs inside the film when submitted to ambient air. Nevertheless, the first external layer, of about 20 nm thickness, is contaminated by a higher oxygen content. Oxygen reaches up to 10% of the elemental ratio at the surface of the film (XPS oxygen-profile on Figure 2). This is explained by an instantaneous oxidation process when collecting the films out of the reactor and exposing them to air: those are easily oxidized with oxygen as they are plasma polymers, containing reactive radicals after their synthesis.[37-40]

**XPS measurements: N/C depth profiles of the films**

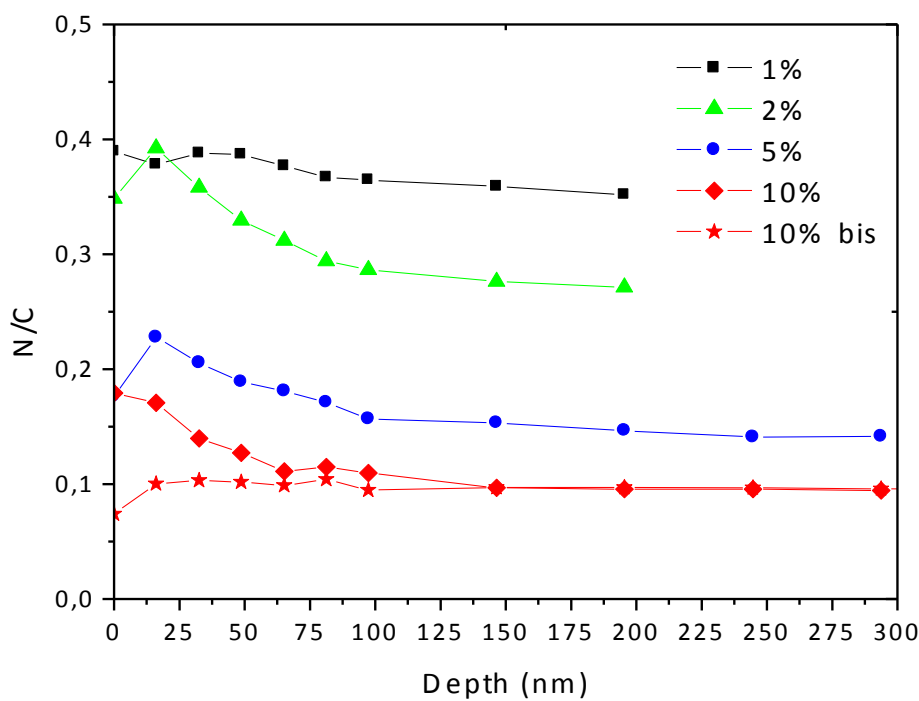

*Figure 3: XPS N/C profiles of tholin films obtained on $SiO_2$ substrates. Two 10% profiles are plotted on both $SiO_2$ and $CaF_2$ (10% bis) substrates.*

We compare the N/C ratios of the films in **Figure 3** obtained by XPS depth profile analysis according to the initial methane concentration in the plasma. The N/C ratio depends strongly on the methane concentration in the plasma discharge. It decreases from about 0.4 for $[CH_4]_0=1\%$, down to 0.1 for $[CH_4]_0=10\%$.

**Comparison with previous XPS measurements on similar materials**

XPS studies were previously performed on other laboratory analogues of Titan's haze. [9, 41, 42] The most resembling material was produced in a spark discharge with a gas mixture of $CH_4/N_2/H_2$.[41, 42] However the material had a strong oxygen signature, also visible through specific shifts in the C1s and N1s XPS spectra. Knowing that a thin oxidized layer is produced on the surface of our films by air exposure, one can suspect a similar contamination in their case, as only the external surface of the material was analyzed by XPS. When no specific treatment is made to protect tholins from air contamination, XPS surface analysis should not be considered as representative of the bulk samples.

**SIMS measurements: no influence of the substrate**

The homogeneity of the films is complementarily probed by SIMS measurements. SIMS provides reliable results after about 100 s of Cesium ion bombardment corresponding to the duration of the SIMS signal stabilization. The first 100 s of film etching cannot be considered in the analysis.

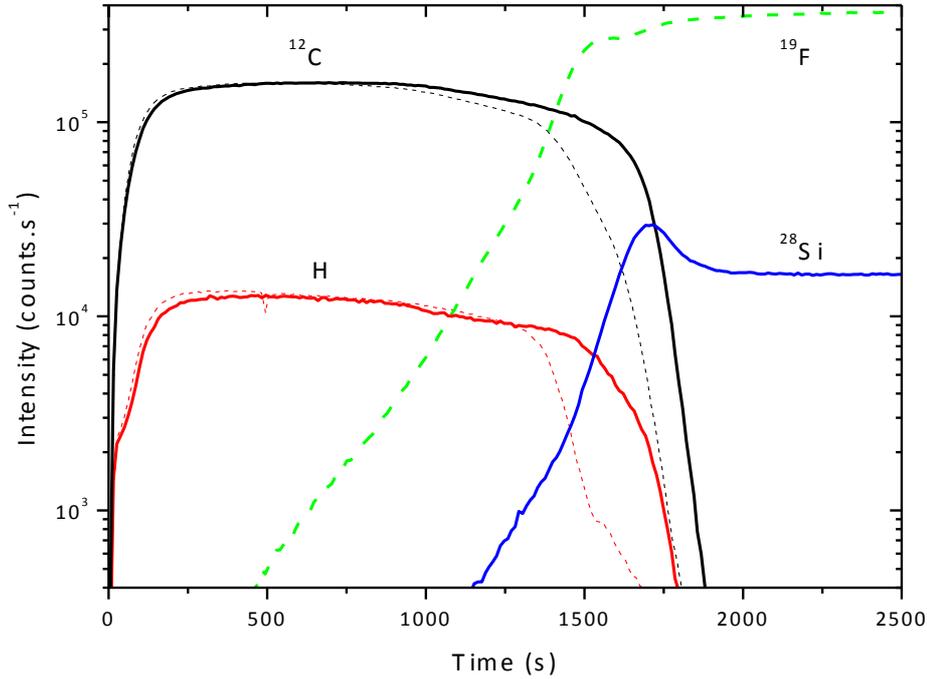

*Figure 4: Comparison of the SIMS profiles obtained in negative mode for thin films produced with a 10% methane gas mixture in similar conditions on respectively a SiO$_2$ (thick plain lines) and a CaF$_2$ substrate (dashed lines). H$^-$ and $^{12}$C$^-$ ion intensities are representative of the film samples, whereas $^{19}$F$^-$ and $^{28}$Si$^-$ ions are respectively representative of the SiO$_2$ and CaF$_2$ substrates.*

The influence of the substrate on the film growth is illustrated on *Figure 4*. This SIMS analysis shows that the samples are homogeneous after a few 100s of SIMS ion bombardment down to the film-substrate interface. SIMS profiles of samples produced with 10% of methane in identical conditions are plotted for films deposited on a SiO$_2$ and a CaF$_2$ substrate. The ions represented are chosen as representative of the organic film ($^{12}$C and H) and of the substrates ($^{28}$Si and $^{19}$F). No significant difference is observed in the main part of the film deposit for the ions representative of the film: the intensities measured for $^{12}$C and H are well reproducible whatever the choice of the substrate. A difference can however be noticed on the film thickness: films are slightly thicker on SiO$_2$ than on CaF$_2$. The initiation of the deposition seems therefore easier on silica substrate, but the resulting films when far enough from the film-substrate interface is not sensitive to this initial discrepancy. For these reasons, films are further analyzed on SiO$_2$ substrates only.

**SIMS measurements: evolution of the films thickness with the initial methane concentration**

The thickness of the films is provided in arbitrary units by the position (sputtering duration) of the sharp decrease observed for carbon ions when Cesium ions encounter the substrate (*Figure 3*4). The decrease occurs at 750s for the 1% film sample, 1000s for the 2% film sample and 1750s for both the 5 and 10% film samples. As all samples have been produced with the same duration of 2 hours, we can conclude that the film deposition rate increases with the initial methane amount in the gas mixture, but reaches an asymptotic value for methane conditions larger than 5%. This evolution is fully consistent with a previous film thickness measurement on Si substrates determined by ellipsometry.[12]

**SIMS measurements: H/C composition of the films**

As explained in the experimental section, the intensity ratio $I_H/I_C$ (*Figure 5*) of the negative ion intensities measured by SIMS is an indicator in arbitrary units of the H/C ratio of the films. We can compare these ratios for the films produced in the different methane amount conditions. $I_H/I_C$ profiles are found to be similar at every profile depths and every CH$_4$ concentrations. The H/C ratio is therefore constant in the films, and insensitive to the methane concentration in the gas mixture.

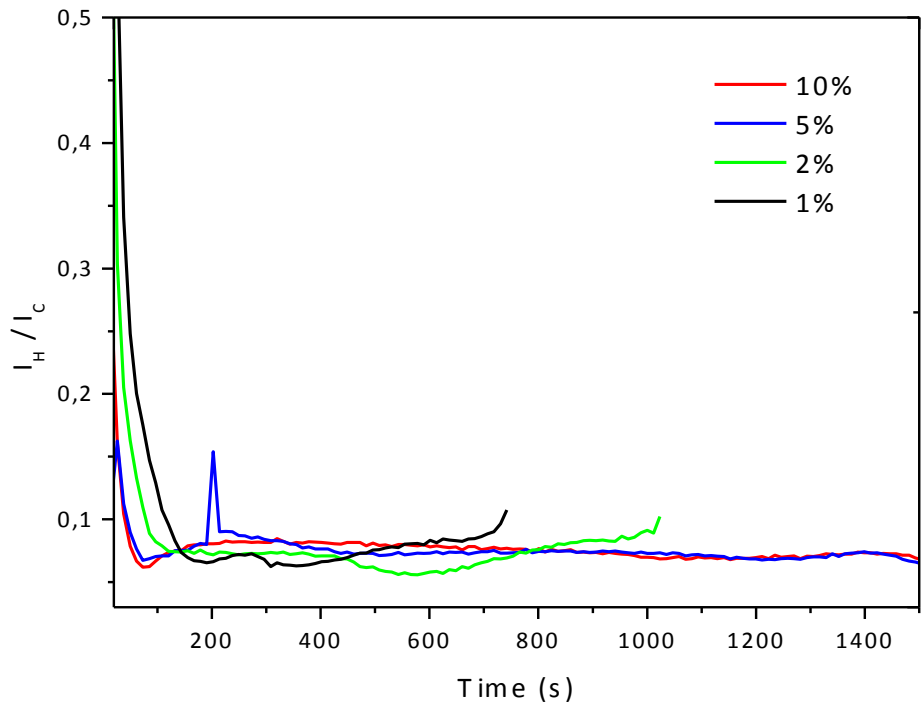

*Figure 5: Influence of the CH$_4$ content on the H/C ratio of the films deduced from SIMS measurements in negative mode (results obtained on SiO$_2$). The I$_H$/I$_C$ ratio is only plotted for the organic film (not beyond the substrate limit).*

Moreover this ratio obtained in negative mode for all the samples can be roughly calibrated thanks to the measurement of the positive molecular ions H-Cs$^+$/ C-Cs$^+$ from the 10% film sample on SiO$_2$ (*Figure 6*). It is found to be equal to about ~0.1. The H/C ratio of the films is thus found to be very low, showing a high unsaturation level in the organic material.

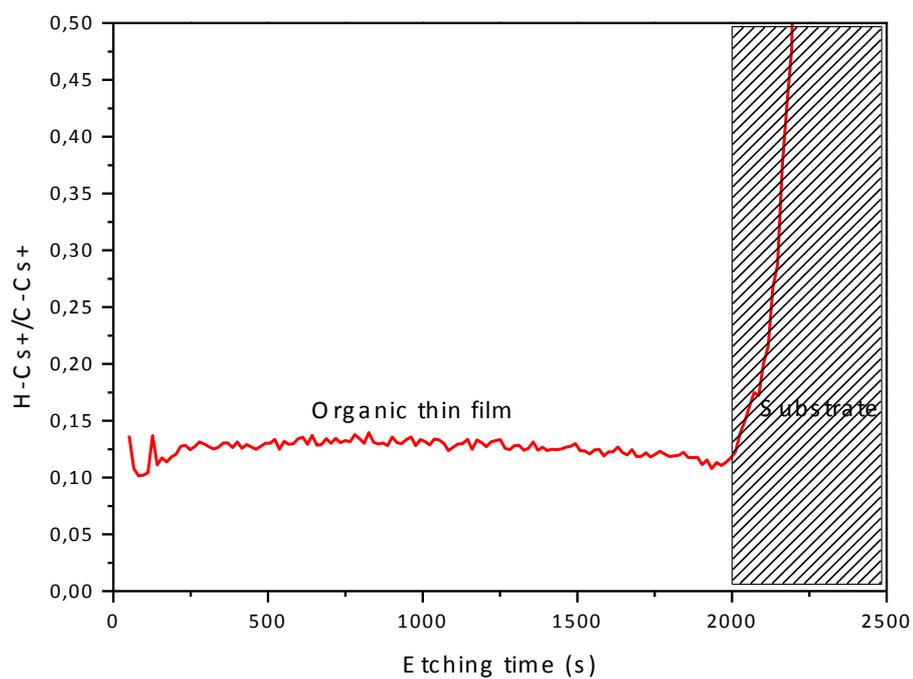

*Figure 6 : Intensity ratio of the molecular ions H-Cs$^+$/ C-Cs$^+$ detected by SIMS on the film sample with a 10% methane gas mixture deposited on a SiO$_2$ substrate.*

## Discussion: comparison with grains in the plasma volume

**Production efficiency**

The film thickness first increases with the initial methane content, from about 500 nm at $CH_4$ 1%, up to about 900 nm at $CH_4$ 5%, with a saturation reached above $CH_4$ 5% [21]. Such an increase, for methane concentrations lower than 5%, is consistent with the production efficiency observed for the grain in similar conditions. As discussed in [20], chemical growth is related to the available quantity of methane dissociated in the discharge, increasing with the methane concentration injected in the reactor.

First, one could expect that films are the results of the accumulation of grains. In this case, the surface roughness of the films should be in the order of magnitude of the grain diameters. This is not the case: films roughness is about a few tens of nanometers whereas grains diameter is about several hundreds of nanometers [12]. Films and grains have therefore different growth modes.

A difference is observed concerning the two solid productions in the plasma at methane concentrations larger than 5%. The film thickness remains maximal (also observed in [12]), whereas the grains almost disappear from the plasma volume (production rate ten times lower at 10% than at 5%) [20]. The limiting factor for the organic growth of the grains at high methane amount is attributed to atomic hydrogen, whose content in the gas phase increases with methane [43]. As no decrease is observed for the films thickness at large methane amounts, this process occurring in the volume is not significant in the case of the films. Other processes are at stake in the plasma limiting the film thickness on the anode.

Films grow during the whole 2 hours experience-duration in the discharge, but their thicknesses do not exceed several hundreds of nanometers whatever the methane concentration [12]. Grains grow in the discharge up to a limited diameter above which they are ejected from the plasma. Their diameters reach the same order of magnitude as the 2-hrs films, whereas they are evacuated from the plasma volume after a few minutes [11]. An important process therefore limits the organic growth of the films on the anode. At the anode, films are exposed to an efficient bombardment of both electrons and negative ions [12, 17]. This bombardment surely erodes the organic films by etching, as observed after argon ion bombardment on carbonitride films [44].

| **Samples** | **Thin films** | | **Grains** | |
|---|---|---|---|---|
| | N/C (XPS) | H/C (SIMS) | N/C | H/C |
| **1%** | 0.4 | ~0.1 | 0.9 | 1.1 |
| **2%** | 0.3 | ~0.1 | 0.8 | 1.1 |
| **5%** | 0.2 | ~0.1 | 0.6 | 1.2 |
| **10%** | 0.1 | ~0.1 | 0.4 | 1.4 |

Table 2: Elemental analysis of both thin films and grains. Thin films N/C ratios are determined here from XPS data at a 150 nm depth; and thin films H/C ratios are taken from Cs-M$^+$ molecular ions from SIMS analysis. The grains elemental analysis has been made in [20, 45].

**Chemical composition.**

Elemental analysis of both films and grains is reported in Table 1.

Films are found less rich in nitrogen by a factor of 2 to 3, compared to grains. This lower nitrogen content explains the lower amine signature compared to the aliphatic one observed in a previous first study [16] opposing films and grains by mid-infra-red absorption spectroscopy.

The H/C ratio is constant in the films, which is not the case for the levitating dust particles, as studied in [20, 45] by elemental analysis. H/C ratio in grains was indeed found to increase with the initial methane amount: C was found constant by elemental analysis, meaning that this ratio increase corresponds to a higher H amount in the grains (in agreement with the suspected role of H as inhibitor for grains growth [43]). Moreover, the similar ~0.1 H/C ratio obtained by SIMS for all films is much lower than the H/C observed in the grains. This highlights a higher unsaturation level in the organic film material, which was not detected by a previous comparative study in infrared absorption spectroscopy [16]. This study actually noted a more intense signature of aliphatic $CH_2/CH_3$ functions in the films than in the grains for a same methane amount in the gas phase, concluding on possible higher hydrogen content in the films. We show here that the hydrogen content is on the contrary globally lower in the films than in the grains. The aliphatic enhancement noticed in [16] is actually explained by a relative decrease of the adjacent amine signature in the films compared to the grains.

As discussed above, the films are located on the grounded electrode during the whole duration of the discharge and are subject to an etching process from the electrons and the negative ions of the discharge. Our work suggests that the grains are a more pristine organic material, and the films could be the result of an additional efficient plasma etching.

Electron etching is known to induce dehydrogenation on carbonitride films [46], in agreement with the low H/C ratios of the films. Moreover a study of argon ion implantation on silicon carbonitride films also observed a decrease of the nitrogen content of the films after ion bombardment, in agreement with the low N/C ratios of the films compared to the grains [44]. These two H and N losses from the films during plasma etching suggest a release of nitrogen-bearing volatiles in the discharge. This possible release is in agreement with the production of ammonia $NH_3$, detected previously in the discharge and hardly explained by only gas-phase chemistry [43, 47].

## Conclusion

The chemical composition of thin films deposited on the grounded anode, in a $N_2$-$CH_4$ RF-plasma discharge is studied by complementary XPS and SIMS analysis. The main plasma parameter studied here is the initial methane amount in the reactive plasma. Films are found to be homogeneous along their depth profile, except an external oxidized layer no thicker than 20 nm and due to the samples air exposure. These films are compared to another solid phase produced simultaneously as spherical grains of about a few hundreds nm diameter in suspension in the plasma volume. Significant differences are found between these two materials: films are more cross-linked and less rich in nitrogen than the grains.

These differences could be explained by additional etching processes on the films through electron, and negative ion bombardment.

In a previous study we measured the optical indices of the films according to the methane concentration in the discharge. The imaginary optical index k, characterizing the absorption properties of the sample, was found to increase drastically with the nitrogen content in the films (by an order of magnitude between films produced at 10 and 2% of methane respectively) [12]. As our analysis reveals that grains are much richer in nitrogen than films for similar methane conditions, we expect that grains are more absorbent than films produced in similar experimental conditions. Hence due to the absence of wall effect during their synthesis, grains are often considered as more representative of Titan's atmospheric aerosols. It would therefore be important to complementarily measure the optical indices of the grains in order to improve the predictions of Titan's radiative models.

**Acknowledgements:** NC acknowledges the financial support of the European Research Council (ERC Starting Grant PRIMCHEM, grant agreement n°636829).